\documentclass[num-refs]{wiley-article}

\usepackage{siunitx}
\usepackage{lineno,hyperref}
\usepackage{graphicx}
\usepackage{slashbox,pict2e}
\usepackage{setspace} 
\usepackage [skip=10pt plus1pt]{parskip}
\usepackage{caption}
\usepackage{subcaption}
\usepackage{multirow}

\graphicspath{ {./images/} }

\papertype{Original Article}
\paperfield{Advanced Sustainable Systems}

\title{Data-Driven Cooling Tower Optimization: A Comprehensive Analysis of Energy Savings using Microsand Filtration}

\author[1\authfn{1}]{Xavier Lefebvre}
\author[2]{Vaishali Ashok}
\author[2]{Dominique Claveau-Mallet}
\author[1]{Etienne Robert}
\author[2]{Emilie Bedard}

\contrib[\authfn{1}]{Corresponding author.}

\affil[1]{Department of Mechanical Engineering, Polytechnique Montreal. 2500 Chem. de Polytechnique, Montreal, QC H3T 1J4}
\affil[2]{Department of Civil Engineering, Polytechnique Montreal}

\corraddress{Department of Mechanical Engineering, Polytechnique Montreal. 2500 Chem. de Polytechnique, Montreal, QC H3T 1J4}
\corremail{xavier.lefebvre@polymtl.ca}

\runningauthor{Lefebvre et al.}

\begin{document}

\begin{frontmatter}
\maketitle

\begin{abstract}
\footnotesize Effective management of cooling tower systems requires thorough disinfection. While traditional chemical water treatment methods are currently the most prominent strategy, they are costly and yield limited results when relied upon as the sole approach. Cross-flow microsand filtration systems offer a promising alternative with the added benefit of potentially increasing evaporative cooling efficiency, thus saving energy. A comprehensive data-driven analysis over two cooling seasons evaluated the energetic performance of a system equipped with and without an operating filter. For similar environmental conditions, the coefficient of performance was on average 18\% higher and was higher 63\% of the time when the filter was operating, indicating superior heat transfer efficiency and significant energy savings. It was also 41\% higher during periods of high cooling demand. Consequently, the filter and the system work more efficiently at high wet-bulb temperature and thermal load. Machine learning modeling suggested that operating the filter year-round could save between 5\% and 13\% of the energy bill, primarily during the cooling season. Continuous filter operation is essential as it mitigates biofouling, underscoring its long-term significance, even during periods of lower thermal loads. Integrating filtration systems into cooling tower management therefore fosters sustainable practices by decreasing energy consumption and biofouling.

\end{abstract}
\end{frontmatter}

\newpage

\section{Introduction}
Operating on the principle of evaporative heat transfer, cooling towers expose warm process water to ambient air, efficiently lowering its temperature \cite{hill1990cooling}. The evaporative cooling efficiency, or performance, can be defined as the ratio of the energy to be evacuated by the tower (thermal load) to the energy required to operate the system \cite{levenspiel2013engineering}. This metric is called the coefficient of performance (COP) and is a measure of the effectiveness of transferring heat between two fluids \cite{linnhoff1979understanding}. Understanding the complex interplay of factors influencing cooling tower performance is essential for optimizing energy efficiency and reducing operational costs. Factors such as the wet-bulb temperature, the lowest atmospheric temperature achievable by evaporating water into a volume of air at constant pressure, as well as water quality and control strategies, all contribute to the overall efficiency of cooling tower operation \cite{baker1984cooling}. Additionally, heat inertia, a measure of heat gradually stored and released in a building, also plays a significant role in determining operational efficiency \cite{verbeke2018thermal}. However, despite technical advancements in cooling tower design, most systems still operate below their optimal efficiency levels \cite{jagadeesh2013performance}. This inefficiency results in increased energy consumption, higher operating costs, and potentially negative environmental impacts \cite{schulze2019life, schulze2018environmental, saidur2010study}. As such, there is a strong need to address these inefficiencies through the integration of innovative technologies. 
\par
Cooling tower systems are not only comprised of the tower itself but also include pumps, heat exchangers and chillers. These elements are typically connected via a piping network, offering a large surface area for biofilms to develop and thrive \cite{paniagua2020impact}. Biofilm is a structured community of microorganisms attached to the surfaces, and its accumulation can act as an isulating layer, decreasing the heat transfer rate by up to three orders of magnitude \cite{habbart2009treatment}. Biofilms can also reduce the effective diameter of pipes, impeding the flow of water and reducing the heat transfer efficiency as less water is exposed to the surrounding air \cite{microorganisms9030577}. Thorough removal of biofilm is particularly difficult due to its complex structure and the presence of protective extracellular polymeric substances, which shield embedded microorganisms from antimicrobial agents, thus necessitating comprehensive strategies for effective control \cite{flemming2010biofilm}. Furthermore, the large surface present in cooling tower systems makes it even more challenging to efficiently disinfect contaminated cooling towers \cite{10.1371/journal.pone.0199429}.
\par
Typical cooling tower water disinfection and conditioning strategies can cost from 50 000 to 300 000 USD annually \cite{cutler2019alternative}. These strategies are often damaging to the cooling tower system over time, inducing significant corrosion and scale formation, which can reduce the heat transfer efficiency by up to two orders of magnitude, and compromise cooling tower performance, potentially causing operational issues \cite{CORTINOVIS20092200, shea1991comparative}. Thus, sustainable methods such as mechanical filtration of the process water have become more relevant in recent years. Cross-flow microsand filtration effectively removes particulate matter and microorganisms larger than 1 $\mu m$ from water \cite{udwin1974side,daalach2017etude}. These systems are usually combined with a lower dosage of the chemical treatment. This approach not only enhances microbial control but also contributes to a reduction of corrosion and scale formation \cite{chien2012pilot,duan2012side}, thereby improving the overall physical state of cooling tower systems. Given the potential of this technology in terms of water treatment, an important question therefore needs to be investigated and quantified: Can the process of filtering the cooling tower water with a microsand filter increase the coefficient of performance of the system?
\par
Cooling tower operation and management typically display significant variability over timescales ranging from minutes to months \cite{10.1371/journal.pone.0199429}. This highlights the importance of continuously monitoring operating cooling towers, which can provide information on the validity of the treatment strategy, as well as the effect of external factors such as thermal load on the operation of the system \cite{seasonCT, etaCT}. However, because of the technical challenges involved in the longitudinal monitoring of cooling towers, such studies are scarce in the literature. Furthermore, limited research has been conducted to explore the impact of filtration on the operational efficiency of cooling towers. The majority of studies have been either analytical \cite{ baker1984cooling, shryock1961comprehensive} or conducted experimentally at reduced size and time scales, rather than being performed \emph{in situ} \cite{bernier1994cooling}. To address this research need, a data-driven comprehensive analysis was conducted on the performance of a cooling tower system equipped with a cross-flow microsand filtering system over 18 months to quantify the effect of this approach on the energy performance. The analysis was carried out on the basis of a system operating alternatively with and without a filter.

\section{Methodology}
A large-scale data acquisition campaign was conducted over two cooling seasons (2022-2023) on an operating cooling tower system. Data from May to September 2022 and from March to September 2023 was considered, as these months are representative of the typical cooling season, associated with high thermal loads \cite{jagadeesh2013performance}. 

\subsection{Site description}
\begin{figure*}[!h]
\makebox[\textwidth][c]{
\includegraphics [width=0.8\textwidth]{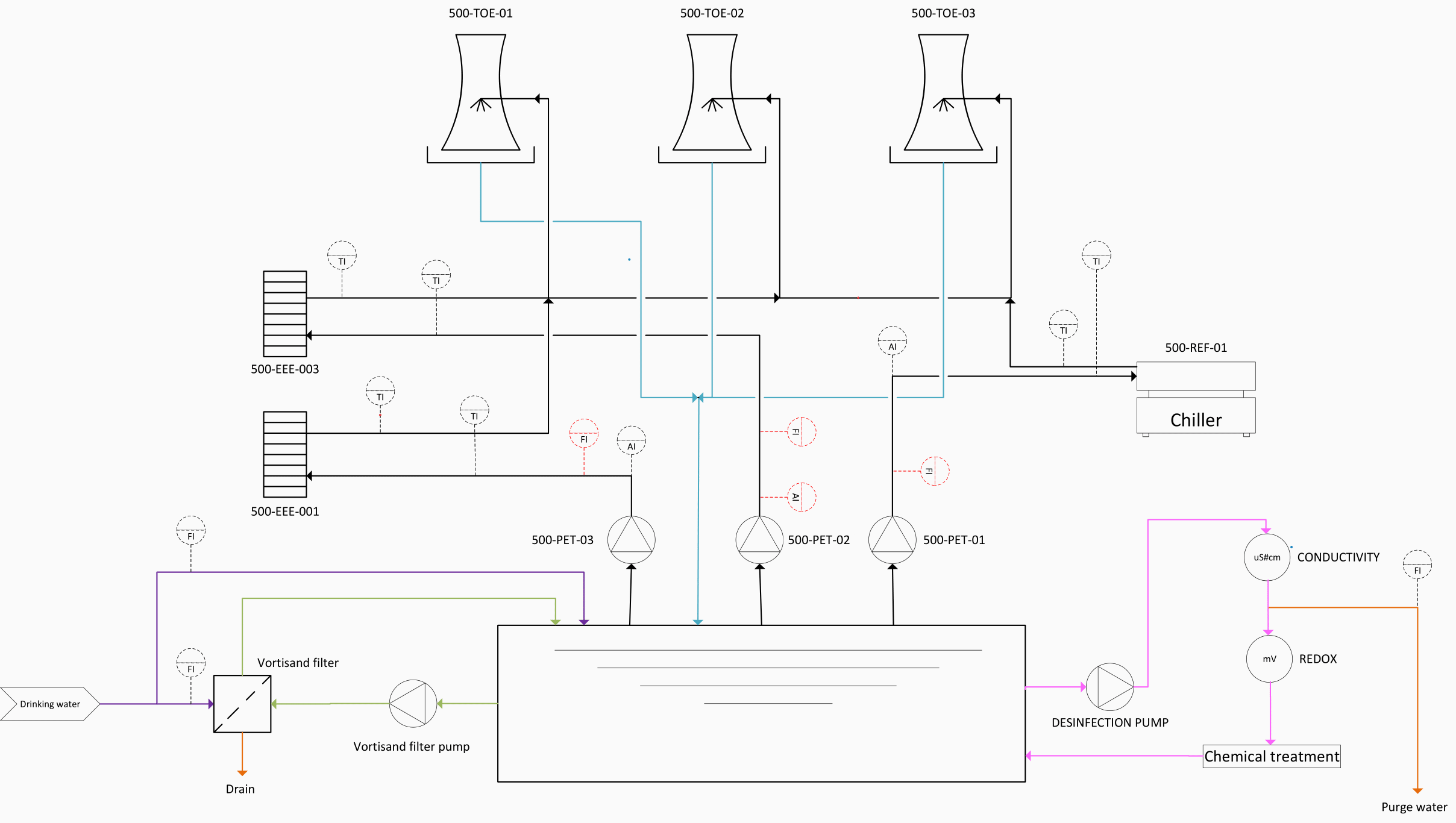}
}
\centering
\caption{Schematics of the cooling tower system and the position of the sensors. Tl are the temperature sensors, Fl the flowmeters and Al the current measurement.}
\label{fig:System}
\end{figure*}

The cooling tower system investigated is implemented in a large building in Montreal, Canada. It can be considered as a large-size system in terms of power consumption \cite{quebec2023comparison}. It is composed of three standard 400-ton (1400 kW) counterflow cooling towers (500-TOE-01, 500-TOE-02 and 500-TOE-03), as well as subsystems including two plate heat exchangers (500-EEE-001 and 500-EEE-003) and one water-cooled chiller (500-REF-01) (Figure \ref{fig:System}). Three vertical centrifugal pumps (500-PET-01, 500-PET-02 and 500-PET-03) feed the water from a basin to each of the subsystems. The water is sprayed inside the cooling towers while ambient air is drawn in with a pair of 10 HP (7.457 kW) axial fans for each cooling tower (not shown in the figure). During the heat transfer, the process water percolates inside the cooling towers and is drained back into a common water basin. Each of the three cooling towers operates independently, based on the total thermal load of the building. Standard chemical treatment is configured and was operated throughout the investigation, with part of the water continuously pumped from the basin to the chemical station and recirculated.
\par

\begin{figure*}[!h]
\makebox[\textwidth][c]{
\includegraphics[width=0.8\textwidth] {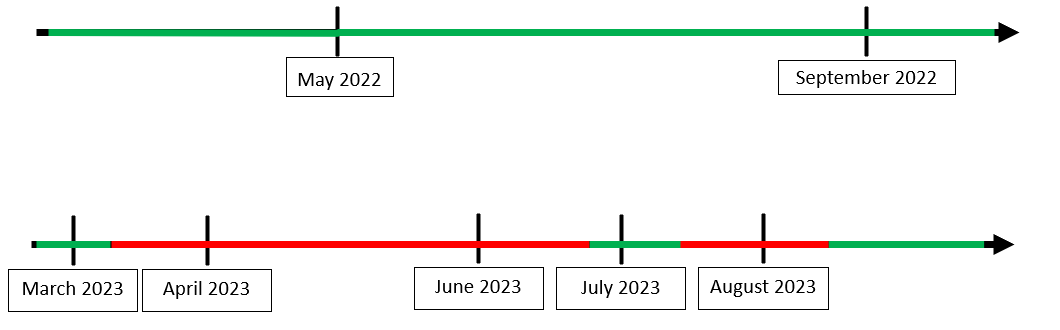}
}
\centering
\caption{Timeline of the Vortisand filtration system operation with regards to the data acquisition campaign. The green line represents the configuration where the filter was operating and the red line configurations where the filter was not operating.}
\label{fig:Timeline}
\end{figure*}

Approximately 20\% of the process water is also circulated through a cross-flow granular microsand industrial filtration system (VORTISAND® Crossflow Microsand Filter) from Xylem+Evoqua. The particularity of this filter is that it generates a cross-flow water pattern that holds the particles in suspension so as not to obstruct the filtering media. This allows the use of a filtering media composed of fine 0.15 mm microsand, resulting in very efficient capture of suspended particles. The filter was initally installed in 2014 and has been running since, except for the intentional breaks in operation for this study. The original installation was done in 2009. The installation date of 2014 refers to the new patented crossflow injectors and media configuration that was installed for the first time insitu for final proofing. The Vortisand filter was operational for the period between May 2022 to March 6th, 2023, and was first shutdown on  March 6th, 2023. It was reactivated from June 26th to July 11th, 2023, then shutdown between July 11th and August 10th, 2023 (Figure \ref{fig:Timeline}). Between May and September 2022, when the filter was always operating, only heat exchanger 1 (500-EEE-01) was operated. Between March and September 2023 however, heat exchanger 2 (500-EEE-03) was also operated.

\subsection{Data Collection}
The dataset was acquired through strategically positioned sensors in the cooling tower system (Figure \ref{fig:System}). To capture the mass flow rate, three flow meters were positioned downstream of each pump (Fl). The temperature profiles within the subsystems were recorded using six sensors (Tl) positioned both downstream and upstream from each heat exchanger and three more temperature sensors monitored the water temperature inside each cooling tower (not shown in figure). The ambient wet-bulb temperature and pressure were also collected from the meteorological data of Environment Canada \cite{noauthor_historical_nodate}. Additionally, the electric current consumption for each pump (Al) as well as the operational speed and nominal power consumption of the cooling tower fans were recorded. Finally, the duty cycles of each component of the system were deduced from the power consumption data. Datapoints for the temperature were acquired at regular 5-minute intervals throughout the two years of the data acquisition campaign. Datapoints for the mass flow rate were acquired at regular 30-minute intervals and assumed to be constant for 5-minute subintervals. 

\subsection{Thermodynamic energy balance}
\begin{figure*}[bt]
\centering
\begin{subfigure}[bt]{0.75\textwidth}
\centering
\includegraphics [width=0.7\textwidth]{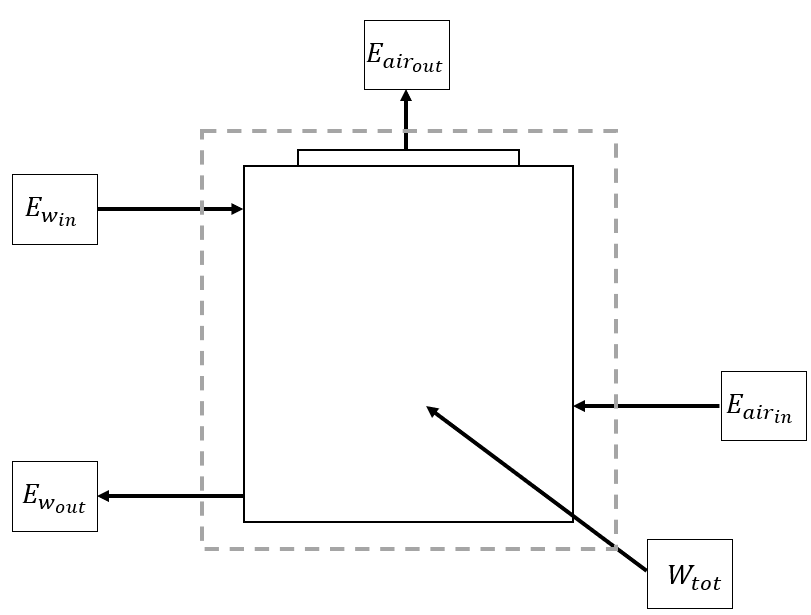}
\caption{}
\end{subfigure}
\begin{subfigure}[bt]{0.6\textwidth}
\centering
\includegraphics[width=0.4\textwidth] {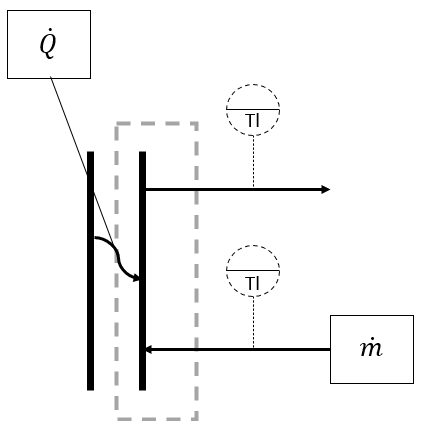}
\caption{}
\end{subfigure}
\caption[Energy assessment]{ a) Thermodynamic energy balance conducted on cooling tower 1. The dotted square represent the system on which the assessment is conducted to define the energy that goes in and out. ${E}_w$ and ${E}_{air}$ represent the energy carried by the water and the air respectively inside and outside the system. The work required to draw the air and the water into the system is given by ${W}_{tot}$. b) Schematics of heat exchanger 1 (500-EEE-01) and the position of the temperature sensors. $\dot{Q}$ represents the heat absorbed into the cooling tower system from the building (thermal load), $\dot{m}$ the mass flowrate of the process water and Tl the temperature of the water before and after the heat transfer. Only the side related to the cooling tower system was considered.}
\label{fig:System_Thermo}
\end{figure*}

To assess the effect of filtration on the performance of the system, a thermodynamic energy balance on the whole cooling tower system was conducted (Figure \ref{fig:System_Thermo} a). Because the cooling towers are interconnected and part of the same system, they can thermodynamically be considered as a singular entity rather than individual units. For each of the 5-minute intervals of data recorded, the system was assumed to be in steady state, the heat transfer process was considered isobaric and possible changes in kinetic and potential energy were neglected, the specific heat of water was considered constant. Finally, the heat released into the environment by the cooling tower was considered equivalent to the heat exchanged in the subsystems. 
\par

The principle of evaporative cooling involves heat transfer between the cooling tower water and the colder ambient air circulating through the system. From the first law of thermodynamics, the energy balance,can be expressed as a function of the energy carried by the water ${E}_w$ and by the air ${E}_{air}$ as well as the work required to draw the air and the water into the system ${W}_{tot}$ (Equation \ref{eq:bilan}): 
\begin{equation*}
{E}_{in} = {E}_{out}
\end{equation*}
\begin{equation}
{E}_{w_{in}} + {E}_{{air}_{in}} + {W}_{tot} = {E}_{w_{out}} + {E}_{{air}_{out}}
\label{eq:bilan}
\end{equation}
From this energy balance, two important parameters were defined: ${E}_{co} = {W}_{tot}$, the electric energy consumed to operate the system and ${E}_{ex}$, the thermal energy exchanged inside the system, which is the energy associated with the thermal load (Equation \ref{eq:Eex}). 
\begin{equation}
{E}_{ex} = {E}_{w_{in}} - {E}_{w_{out}}
\label{eq:Eex}
\end{equation}
These two parameters were used to assess the coefficient of performance of the system and thus, the efficiency of the system at conducting heat exchange with and without the filter operating. 

\subsubsection{Energy performance parameter}
A modified coefficient of performance (COP), was applied to the whole cooling tower system instead of the typical definition for subsystems (Equation \ref{eq:COP}): 

\begin{equation}
COP = \frac{{E}_{ex}}{{E}_{co}}
\label{eq:COP}
\end{equation}

\subsubsection{Electric energy consumed by the system}
${E}_{co}$ was calculated from the electric energy consumed by each pump and fan, as well as by the pump of the filter. 
\par
The electric energy consumed by the pumps was calculated from their power consumption $P_{pump}$ (W), knowing the constant working voltage $V$ (V) of 575 V and the measured operating electric current $I$ (A). 
\begin{equation}
P_{pump} = V I
\label{eq:P_pump}
\end{equation}
The electric energy consumed by the pairs of cooling tower fans was calculated from their power consumption $P_{fan}$ (kW) using the fan affinity law for power \cite{heald1988cameron}, with knowledge of the fan operating speed $N_{fan}$ (RPM) and the nominal power consumption $P_0$ (kW), which was of 8.3375 kW when the fan is operating at a nominal speed $N_0$ of 60 RPM. 
\begin{equation}
P_{fan} = P_0 \left(\dfrac{N_{fan}}{N_0}\right)^{3}
\label{eq:P_fan}
\end{equation}
\par
From the power consumption of the pumps and the pairs of fan, the respective energy consumed (kWh) was calculated by multiplying the power consumption $P$ (kW) of each of the elements with the time of use $t$ (h). The electric energy consumed by the Vortisand filter ${E}_{Vortisand}$ was constant at 0.31 kW. Finally, the total electric energy consumed by the system ${E}_{co}$ over time $t$ was given by the sum of the energy consumed by each of the elements.
\begin{equation}
{E}_{co} = \sum P t
\label{eq:E_chaleur}
\end{equation}
\begin{equation*}
{E}_{co} = {E}_{Pump1}+{E}_{Pump2}+{E}_{Pump3}+{E}_{Fan1}+{E}_{Fan2}+{E}_{Fan3}+{E}_{Vortisand}
\label{eq:E_consumed}
\end{equation*}

\subsubsection{Thermal energy exchanged inside the cooling towers}
Likewise, the thermal energy exchanged by the cooling towers between the building and the environment ${E}_{ex}$ (kWh) was given by the sum of the energy exchanged by each subsystem (both heat exchangers and the chiller). 
\begin{equation}
{E}_{ex} = E_{HeatExchanger1}+E_{HeatExchanger2}+E_{Chiller}
\label{eq:E_exchanged}
\end{equation}
${E}_{ex}$ was calculated from the heat absorbed from the building, equivalent to the thermal load on the system $\dot{Q}_{load}$, multiplied by the time of use $t$ (h)
\begin{equation}
{E}_{ex} = \sum \dot{Q}_{load} t
\label{eq:E_chaleur}
\end{equation}
\begin{equation}
\dot{Q}_{load} = \dot{Q}_{HeatExchanger1}+\dot{Q}_{HeatExchanger2}+\dot{Q}_{Chiller}
\label{eq:Load}
\end{equation}
The heat absorbed from each subsystem $\dot{Q}$ (J/s) was calculated, knowing the mass flowrate of the process water $\dot{m}$ ($kg/s$), the heat capacity of the water $c_p$ (J/kg K) and its temperature difference from before and after the heat transfer $\Delta T$ (K), measured on either side of the subsystem inside the cooling tower system (Figure \ref{fig:System_Thermo}):
\begin{equation}
\dot{Q}= \dot{m} c_{\text{p}} \Delta T
\label{eq:chaleur}
\end{equation}



\subsection{Data analysis}
To assess the effect of the filter on the coefficient of performance, two configurations of the cooling tower system were compared: with and without the filter operating (filter ON and OFF). Time series were defined to compare the configurations during intervals where the cooling tower system experienced similar environmental conditions. 
A large dataset with more than 890 000 entries was considered. The data analysis was conducted using the Julia programming language \cite{bezanson2017julia}. 

\subsubsection{Time series}
The wet-bulb temperature being within a narrow range for both configurations was initially identified as a comparison metric to evaluate the coefficients of performance of the system. Three parameters were considered to define the time series: First, $margin$, the root mean square difference between the wet-bulb temperature of both datasets ($^\circ C$), second $range$, the duration of the period considered for the time series (hours) and third, $inertia$ the heat inertia on the building. The assessment of heat inertia was considered by making sure that the slope of the wet-bulb temperature was of the same sign for a defined period (hours), before the start of the time series. An in-house algorithm scanned through the data to locate temporal correspondences between two input vectors, the wet-bulb temperatures for each configuration of the system, based on the $margin$ and $inertia$. The algorithm employed a sliding window approach. It began with a subset of the input vector for the configuration with the filter operating, starting with the first value and tried to find a match in the other input vector, iteratively analyzing subsequences of length $range$ of the input vectors. Then, the algorithm evaluated the next subset, starting at the second value of the input vector, and repeated the process until all values of the dataset were assessed. When two time series were matching, they were appended in another variable. To enhance computational efficiency, parallel processing was leveraged to concurrently explore potential matches. 
\par
A convergence analysis was conducted to select the three parameters defining the time series. Values for each of the parameters were individually progressively increased until the number of time series reached a plateau, indicating optimal parameter configuration to maximize the size of the dataset. The objective was to strategically select parameter values that would maximize the number of time series generated, while still offering a good resolution on the obtained results. Following this optimization process, a carefully chosen set of parameters for $margin$ (0.5 $^\circ C$), $range$ (6 hours), and $inertia$ (6 hours) yielded a total of 28 265 time series after outliers were filtered out. 

\subsubsection{Machine learning}
A machine learning model was defined to predict how much energy would be saved in a given timeframe if the filter was operating throughout (ON), compared to if it was not operating (OFF). The random forest regressor model from the MLJ.jl package in Julia was adopted due to its overfitting mitigation and improved accuracy by using multiple decision trees \cite{biau2016random} (See Appendix A - supplementary material). These hierarchical structures recursively partition the data based on the parameters of interest to make sequential decisions. 
MLJ (Machine Learning Julia) provides a composable framework for building, evaluating, and deploying machine learning models. The Random Forest Regressor model in MLJ was imported from the TensorFlow package of the Python language \cite{joshi2017artificial}. It uses features (parameters of interest selected from the dataset) to predict the value of a target variable after having been trained on part of the data to build the relationship between the features and the target variable \cite{susnjak2015wisdom, biau2016random}.
\par
Using cross-validation principles, the dataset was separated into the two aforementioned configurations for the predictions, when the filter was operating (ON) and when it was not (OFF). A random set of 80\% of the data was used to train the model and the remaining 20\% was used to test it. The target variable adopted was the electric energy consumed by the system ${E}_{co}$. The features used were the thermal load, the water temperature inside each cooling tower, the wet-bulb temperature, the thermal energy exchanged inside the system ${E}_{ex}$, the coefficient of performance, and the configuration of the filter (ON or OFF). All features were scaled from 0 to 1 using a sigmoid function before building the model, and training it \cite{jayalakshmi2011statistical}. The model was then validated on the testing data using the R-squared metric calculated between the observed and predicted values. To optimize performance, key parameters such as the number of trees (n\_estimators =100), maximum tree depth (set to "None"), and maximum number of features (set to "auto") were fine-tuned \cite{joshi2017artificial}. Once fitted accordingly, the model was then applied to the respective data for the years 2022 and 2023. 

\section{Results and Discussion}

\subsection{Time series validation}
As year-on-year comparisons were not possible because key parameters (wet-bulb temperature, thermal load) could change drastically, time series based on the wet-bulb temperature were adopted. However, given the significant variability associated with the dataset, the definition of the time series can substantially impact the obtained results. Consequently, it was necessary to first validate the chosen parameters before delving into assessment of the energetic performance of the system with or without an operational filter. This validation involves a convergence analysis and a subsequent examination of the coefficient of variation across the selected time series, a ratio of the variability to the mean. The objectives were to maximize data resolution and mitigate the bias associated with the dataset, as well as the impact of variability on the outcomes.
\par
A convergence analysis was first undertaken to select the three parameters defining the time series (Figure S1 - supplementary material). The intent was to maximize the number of time series, while reducing the variability of the dataset, thus facilitating the development of correlations between the performance of the system and the environmental conditions. Increasing the $margin$ parameter from 0.4 to approximately 1$^\circ$C resulted in a linear increase of the number of time series up to more than 94 000. Further increase did not change the number of time series significantly. However, a $margin$ parameter larger than 0.5 introduced significant variance in the time series (Figure S2 - Supplementary material). A value of 0.5$^\circ$C was therefore selected for this parameter to minimize the noise in the time series, and yielded more than 40 000 time series. Conversely, decreasing the values beyond 6 hours for the $range$ and $inertia$ parameters reduced the number of series. This value was therefore selected for both parameters. From this convergence analysis, and after filtering out the outliers, maintaining a maximal root mean square difference of 0.5$^\circ$C in wet-bulb temperature between configurations and considering the thermal load over the preceding 6 hours, 28 265 six-hour time series were generated. The extensive dataset therefore provided sufficient information to define a large number of time series to thoroughly investigate the intricate dynamics of the cooling tower system.

\begin{figure*}[h]
\centering
\includegraphics [width=0.75\textwidth] {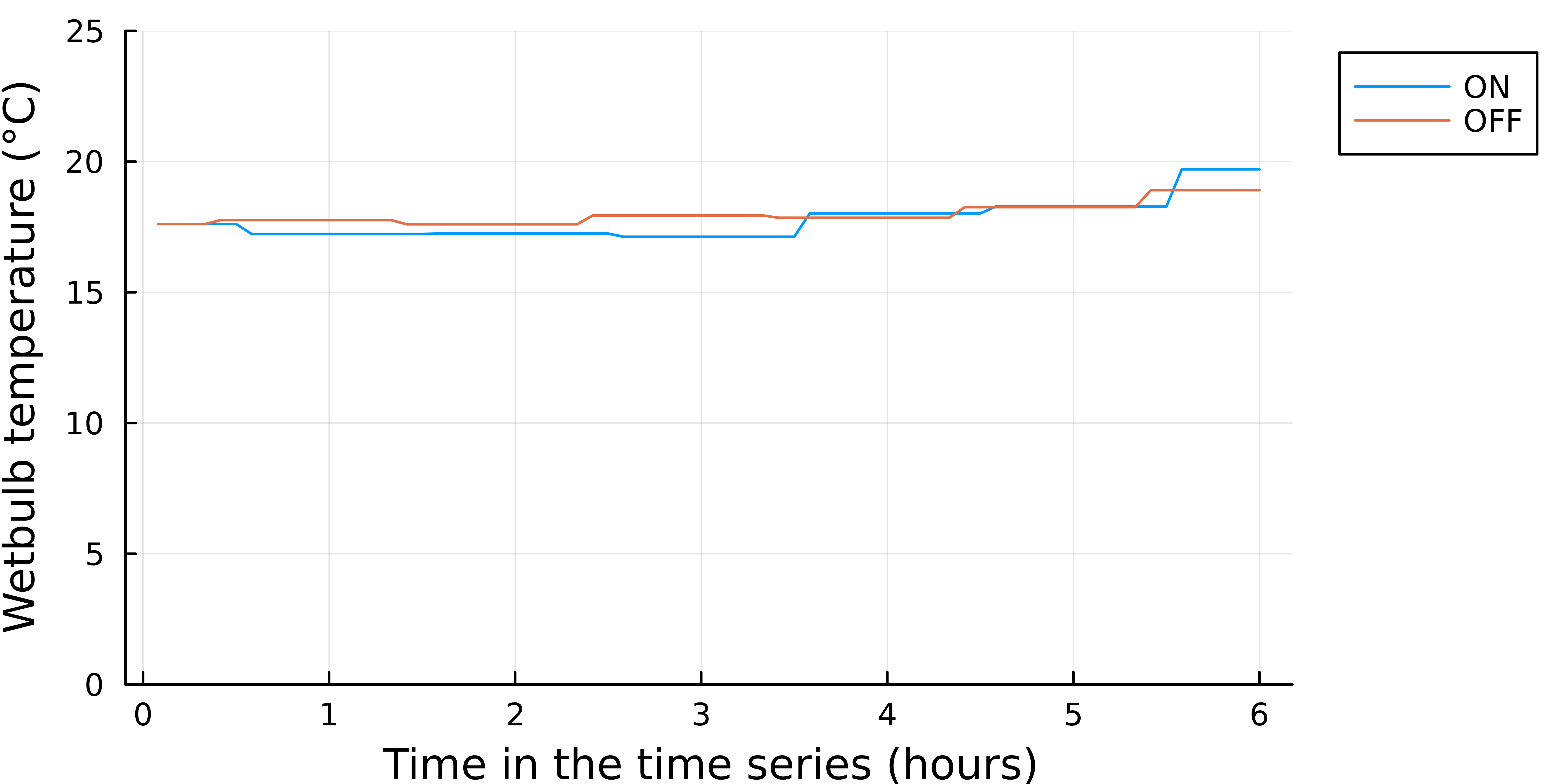}
\caption{Results of the wet-bulb temperature for a signle period of 6 hours. Configurations where the filter was operating (ON) and not operating (OFF) are displayed.}
\label{fig:Timeserie}
\end{figure*}

Wet-bulb temperatures ranging from nearly 0 to 25$^\circ$C were recorded throughout the investigation, with an average coefficient of variation of between 14.6 and 15.4 \% depending on the filter configuration (Table \ref{tab:COV}). The coefficient of variation was used to compare the extent of variability between the individual time series, as well as across all the time series. Figure \ref{fig:Timeserie} a) displays the wet-bulb temperature for a single time series with the selected parameters. As an example, in that specific period of 6 hours, the wet-bulb temperature oscillated between 18 and 20$^\circ$C when the filter was operating, and when it was not. As can be observed, the difference in wet-bulb temperature between both filter configurations is minimal. Furthermore, the difference between the wet-bulb temperature of each configuration represents a coefficient of variation of less than 2\%. This indicates that for each time series, the variability in the wet-bulb temperature was small, but that between all the selected time series, a broad range of wet-bulb temperatures was covered. 
\par

\begin{table}[h!]
\centering
\caption{Coefficient of variation (\%) for individual time series and across all time series for the wet-bulb temperature, for configurations where the filter was and was not operating.}
\begin{tabular}{|c|c|c|c|}
\hline
 &
  \begin{tabular}[c]{@{}c@{}}Configuration \\ of the filter\end{tabular} &
  \begin{tabular}[c]{@{}c@{}}Average coefficient of variation \\ of individual time series\\ (\%)\end{tabular} &
  \begin{tabular}[c]{@{}c@{}}Average coefficient of variation\\ across all time series\\ (\%)\end{tabular} \\ \hline
\multirow{2}{*}{\begin{tabular}[c]{@{}c@{}}Wet-bulb\\ Temperature\\ (C)\end{tabular}} &
  ON &
  1.52 &
  14.58 \\ \cline{2-4} 
 &
  OFF &
  1.78 &
  15.39 \\ \hline
\end{tabular}
\label{tab:COV}
\end{table}

These results highlight the representativeness of the time series for a broad range of environmental conditions during a typical cooling season, reflecting low-bias data selection and further validating the choice of parameters for the time series.

\subsection{Effect of the filtration system on performance}

\begin{figure*}[h]
\centering
\includegraphics [width=0.75\textwidth] {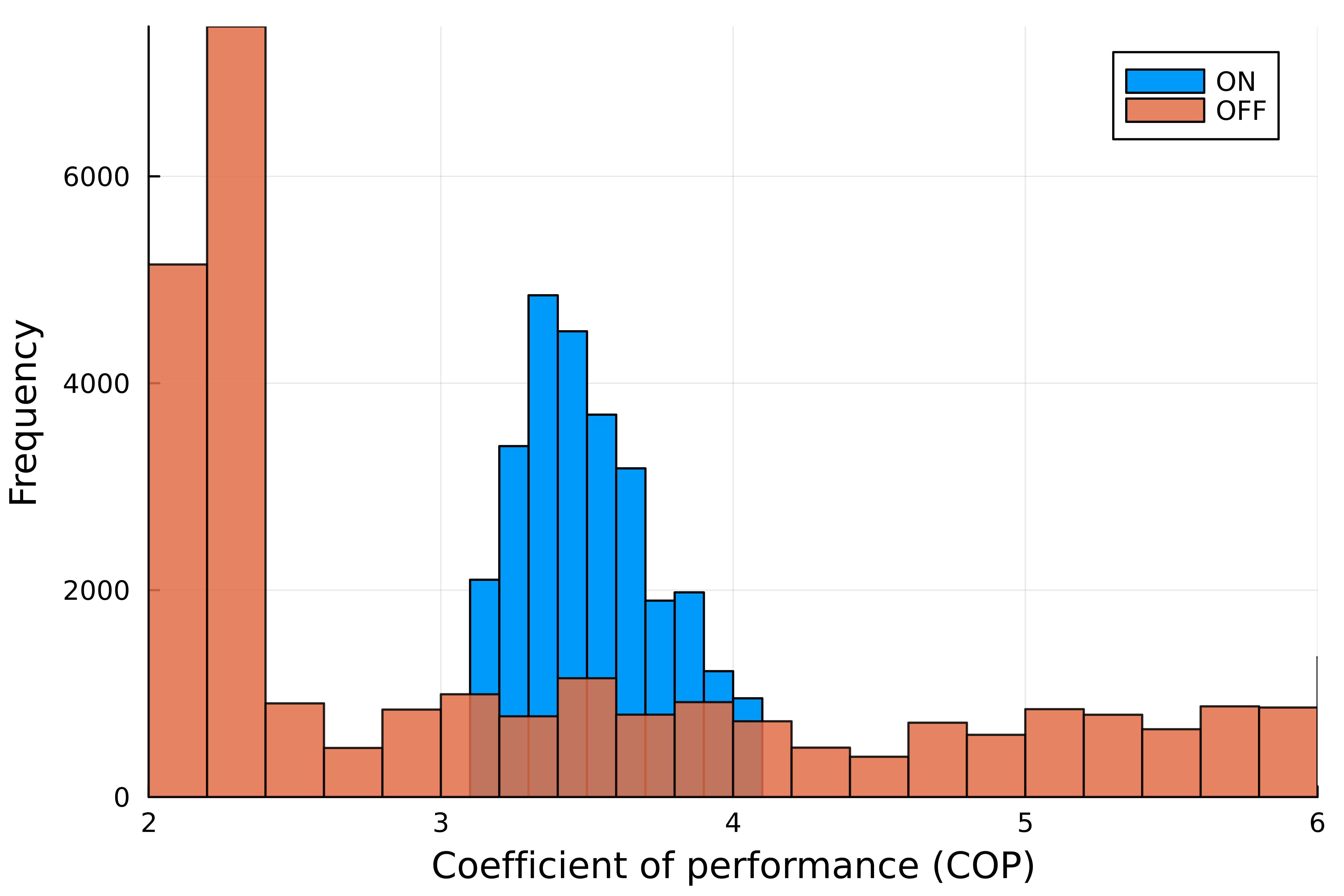}
\caption[COP]{Coefficient of performance of the system across all time series.}
\label{fig:COP}
\end{figure*}

The coefficient of performance was calculated across all the 28 265 time series, and ranged from 2.2 to 6 when the filter was not operating and from 3.2 to 4.1 when the filter was operating, showing more stability for the latter configuration (Figure \ref{fig:COP}). The coefficient of performance, usually calculated for each subsystem, typically ranges between 1 and 5 \cite{riffat2004improving,chen1966correlation}, meaning that the thermal energy transferred inside the system is higher than the mechanical energy provided to the system \cite{davies2004building, incropera1996fundamentals}. The modified metric used in this study, calculating a global COP for the cooling tower system, was slightly higher because of the compounding effect of the multiple subsystems (heat exchangers and chillers). Across all time series, the coefficient of performance was on average 18\% higher and was higher 63\% of the time when the filter was operating, revealing that over time, the cooling tower system was more efficient at conducting heat transfer when the filter was operating. 
\par

However, these results also indicate that under specific conditions, the system was more efficiently conducting heat exchange when the filter was not operating. These performance peaks were associated with periods of low wet-bulb temperature and thermal loads not characteristic of the cooling season. Further investigation was performed to understand conditions affecting the coefficient of performance. 
\par
\begin{figure*}[h]
\centering
\begin{subfigure}[b]{0.7\textwidth}
\centering
\includegraphics [width=0.75\textwidth] {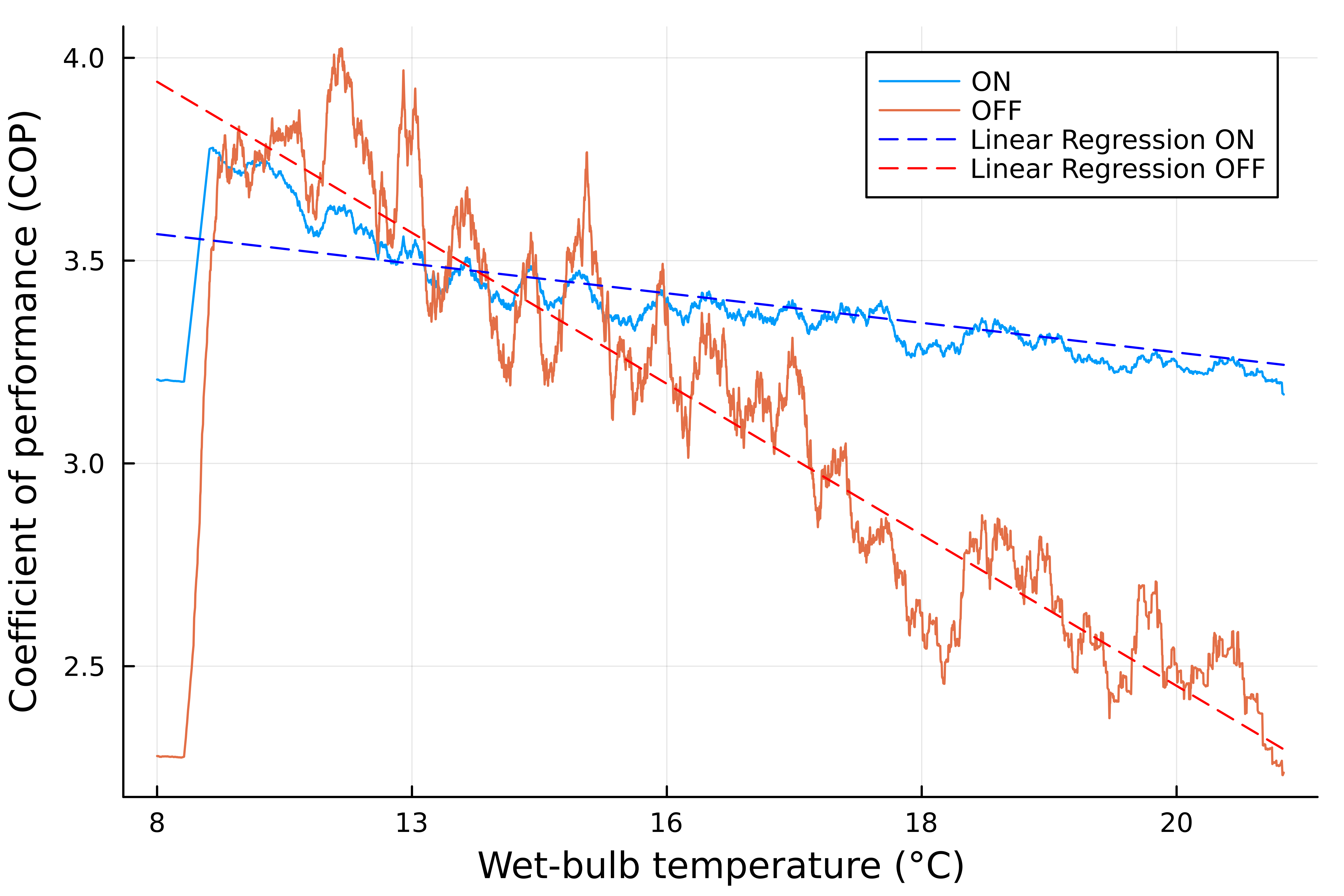}
\caption{}
\end{subfigure}
\begin{subfigure}[b]{0.7\textwidth}
\centering
\includegraphics [width=0.75\textwidth] {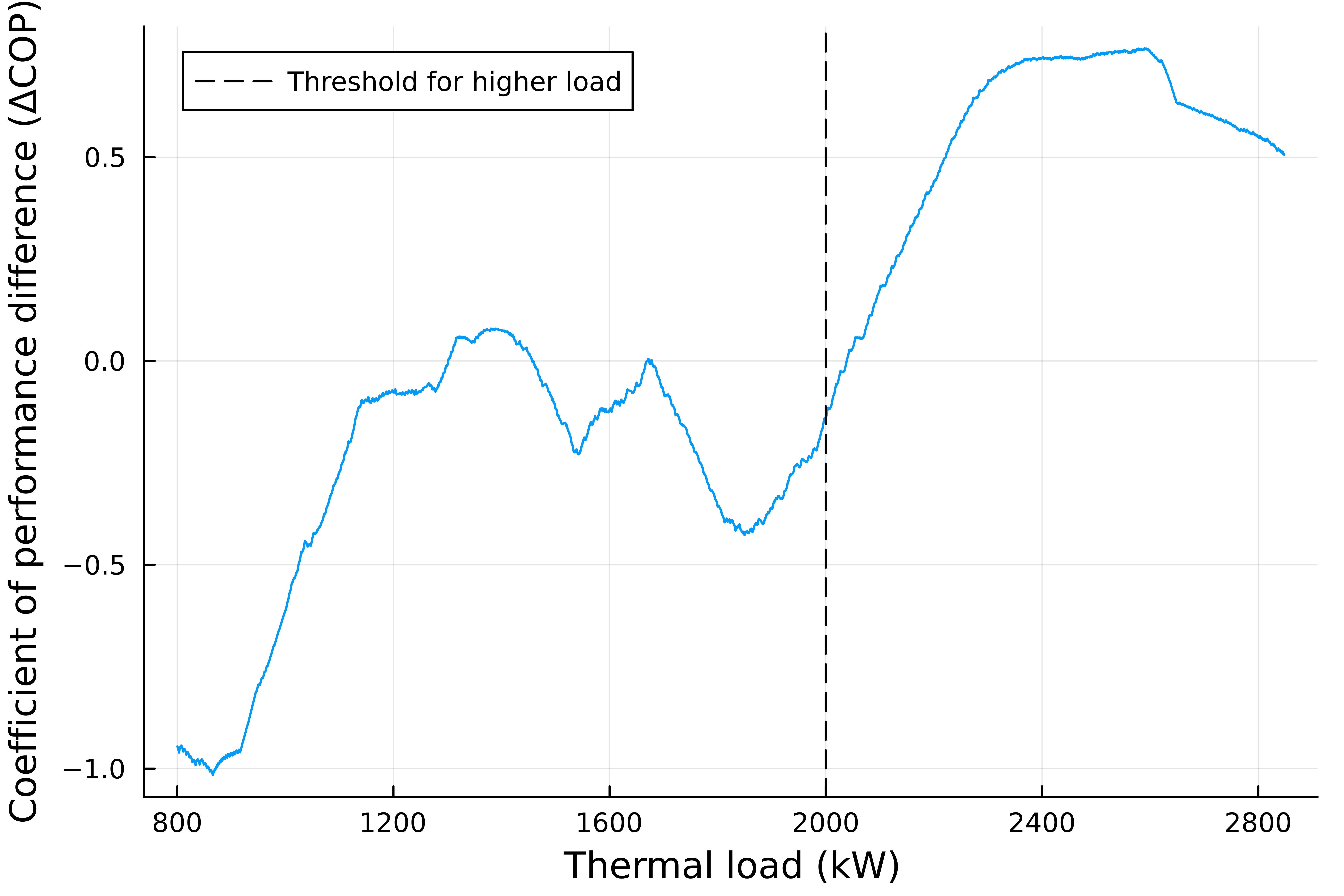}
\caption{}
\end{subfigure}
\caption[COP according to Twb]{a) Coefficient of performance of the system across all time series as a function of the wet-bulb temperature. b) Difference of the coefficient of performance when the filter is ON and when the filter is OFF for wet-bulb temperatures larger than 10 $^\circ$C. Configurations where the filter was operating (ON) and not operating (OFF) are displayed.}
\label{fig:COP_expl}
\end{figure*}

The wet-bulb temperature was found to be a critical factor influencing the coefficient of performance (Figure \ref{fig:COP_expl} a). As can be observed, the coefficient of performance decreased linearly with increasing wet-bulb temperature, likely caused by the combination of the increased load on the system and potential biofilm accumulation (See Appendix C - supplementary material) \cite{al1997optimum}. The slope was steeper (approximately -6\% compared to -2\%) in the configuration where the filter was not operating, indicating that the filter enables the system to work more efficiently at high wet-bulb temperature, associated with the most energy-consuming periods. This phenomenon is of significance for energy conservation because it highlights the potential benefits of implementing strategies to mitigate the impact of higher wet-bulb temperatures on cooling tower behavior and performance.
\par

The thermal load was also significantly correlated to the energy exchanged inside the system. The difference in the coefficient of performance between the configuration with the filter ON and with the filter OFF was found to vary according to the thermal load (Figure \ref{fig:COP_expl} b). A positive difference in the coefficient of performance indicates a higher coefficient of performance when the filter was operating than when it was not. Although not linearly, the difference in the coefficient of performance increased with the thermal load. Instances where the coefficient of performance was larger when the filter was not operating occurred only for thermal loads smaller than approximately 2000 kW, which can be associated with periods of low to medium cooling demand for a large cooling tower such as the investigated system \cite{quebec2023comparison,al1997optimum,balaras1996role}. This indicates that the effect of the filter on the coefficient of performance was more prominent for higher thermal loads. 
\par
\begin{figure*}[h]
\centering
\includegraphics [width=0.75\textwidth] {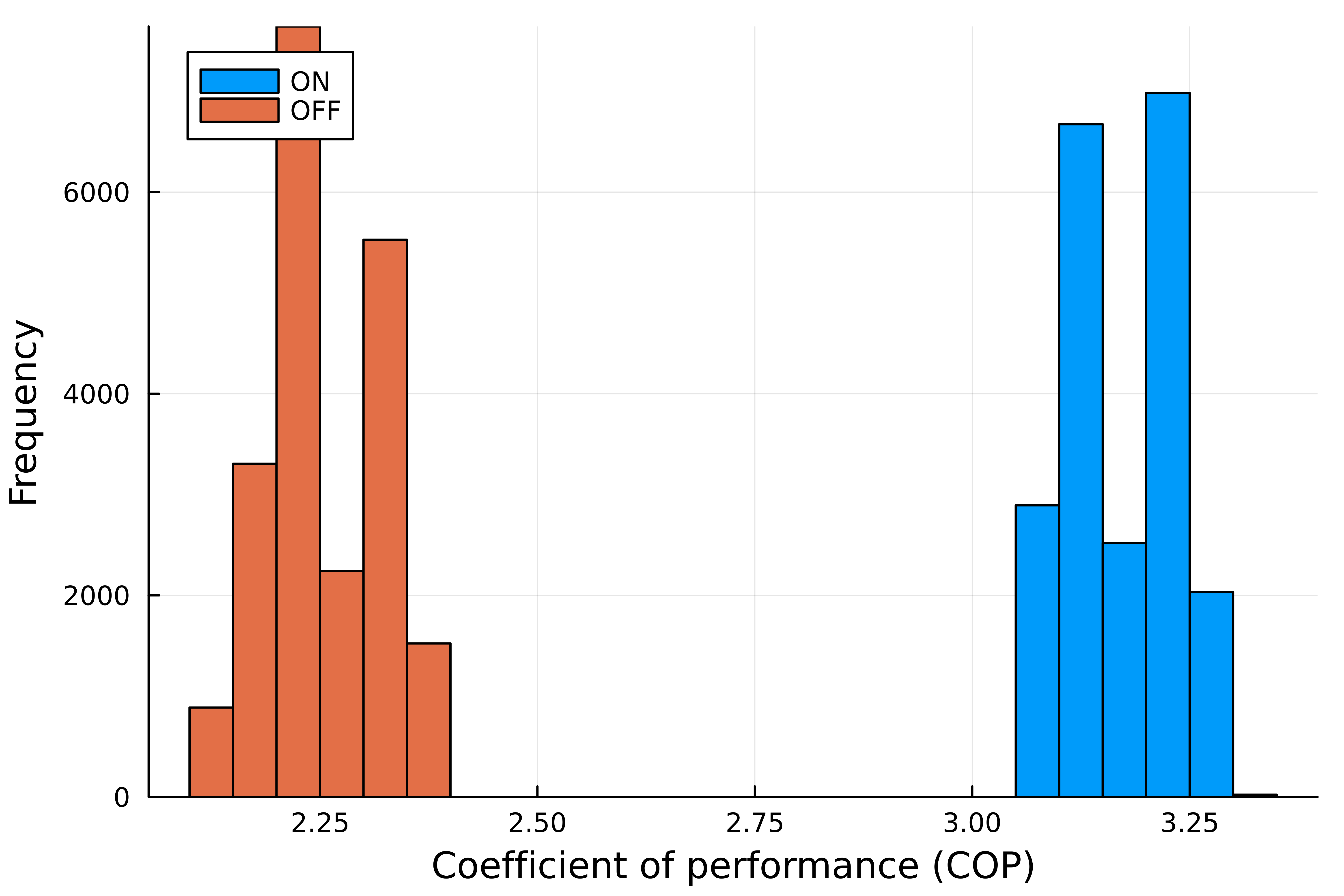}
\caption[COP according to Twb]{Coefficient of performance of the system for wet-bulb temperatures larger than 10$^\circ$C and thermal load larger than 2000 kW. Configurations where the filter was operating (ON) and not operating (OFF) are displayed.}
\label{fig:COP_TwbLoad}
\end{figure*}

When considering only periods of high cooling demand, associated with thermal loads higher than approximately 2000 kW and wet-bulb temperatures higher than 10 $^\circ$C, the coefficient of performance was always higher with the operating filter (Figure \ref{fig:COP_TwbLoad}). More than half of the time series respected these thresholds, which yielded a coefficient of performance 41 \% higher when the filter was operating, as well as a standard deviation of approximately 0.1, indicating high consistency when the cooling demand is high. Thus, the system was found to be more efficient for 81\% of the total transferred energy during the investigation, most markedly during periods of high cooling demand. 
\par

\subsection{Energy Savings Prediction}
This investigation revealed a consistent trend wherein the cooling tower system demonstrated superior heat transfer efficiency over time when employing filtration, irrespective of prevailing conditions. Moreover, higher cooling demands corresponded to an increase in the system efficiency at conducting heat transfer, which was particularly evident when the filter was operational. During periods of high thermal load, the electricity demand for cooling tower operation increases significantly. Consequently, any reduction in electricity consumption, particularly during these peak periods, is not only beneficial in terms of immediate energy savings but also contributes to the overall durability of the system. By mitigating energy demand during peak times, filtration systems offer a sustainable solution for managing cooling tower operations efficiently while reducing strain on the power grid and minimizing the environmental footprint of industrial processes. Consequently, the integration of filtration technologies is promising for long-term energy conservation. However, the precise extent of energy savings attributable to filter usage could not be established for a full year due to  the intermittent activation and deactivation of the filter during the investigation.
\par
\begin{figure*}[!h]
\makebox[\textwidth][c]{
\includegraphics[width=0.75\textwidth] {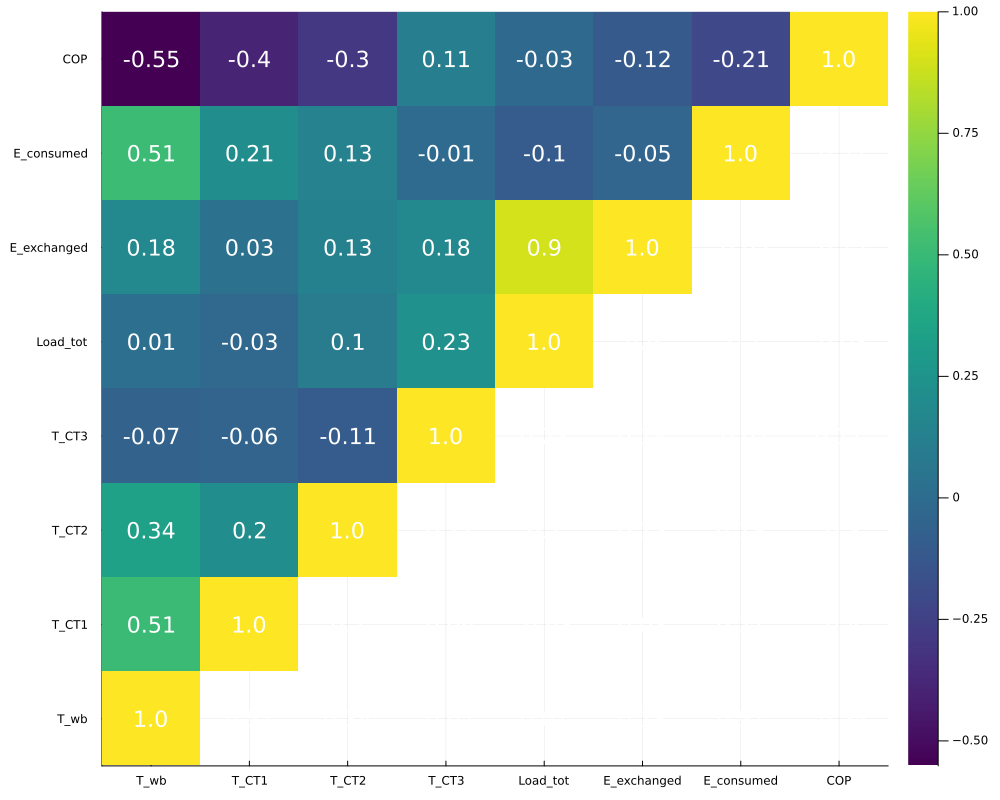}
}
\centering
\caption{Heatmap of the correlations between the different features (the wet-bulb temperature $T_{wb}$, the temperature of the water inside each cooling tower $T_{CT1}$, $T_{CT2}$ and $T_{CT3}$, the thermal load $Load_{tot}$, the thermal energy exchanged $E_{exchanged}$ and the coefficient of performance $COP$) and the target (The electric energy consumed by the system $E_{consumed}$). A positive correlation indicates that when one of the variables increases, the other increases as well. A negative correlation indicates that when one of the variables increases, the other decreases. The higher the value of the correlation the more important the increase or decrease. }
\label{fig:Heatmap}
\end{figure*}

To quantify the potential energy savings associated with specific operational scenarios, a predictive model was developed to estimate the energy consumed throughout the year based on observed external conditions affecting the cooling tower system. 
Figure \ref{fig:Heatmap} displays the strength of the correlation between different environmental parameters within the cooling tower system. This heat map was used to determine which external conditions should be considered to define the model. The highest correlation observed for the electric energy consumed was with the wet-bulb temperature at 0.51, but the other displayed parameters were also significantly correlated to the consumed energy. These parameters were thus selected to develop the machine learning model.
\par

\begin{figure*}[h]
\centering
\begin{subfigure}[b]{0.7\textwidth}
\centering
\includegraphics [width=7.5cm] {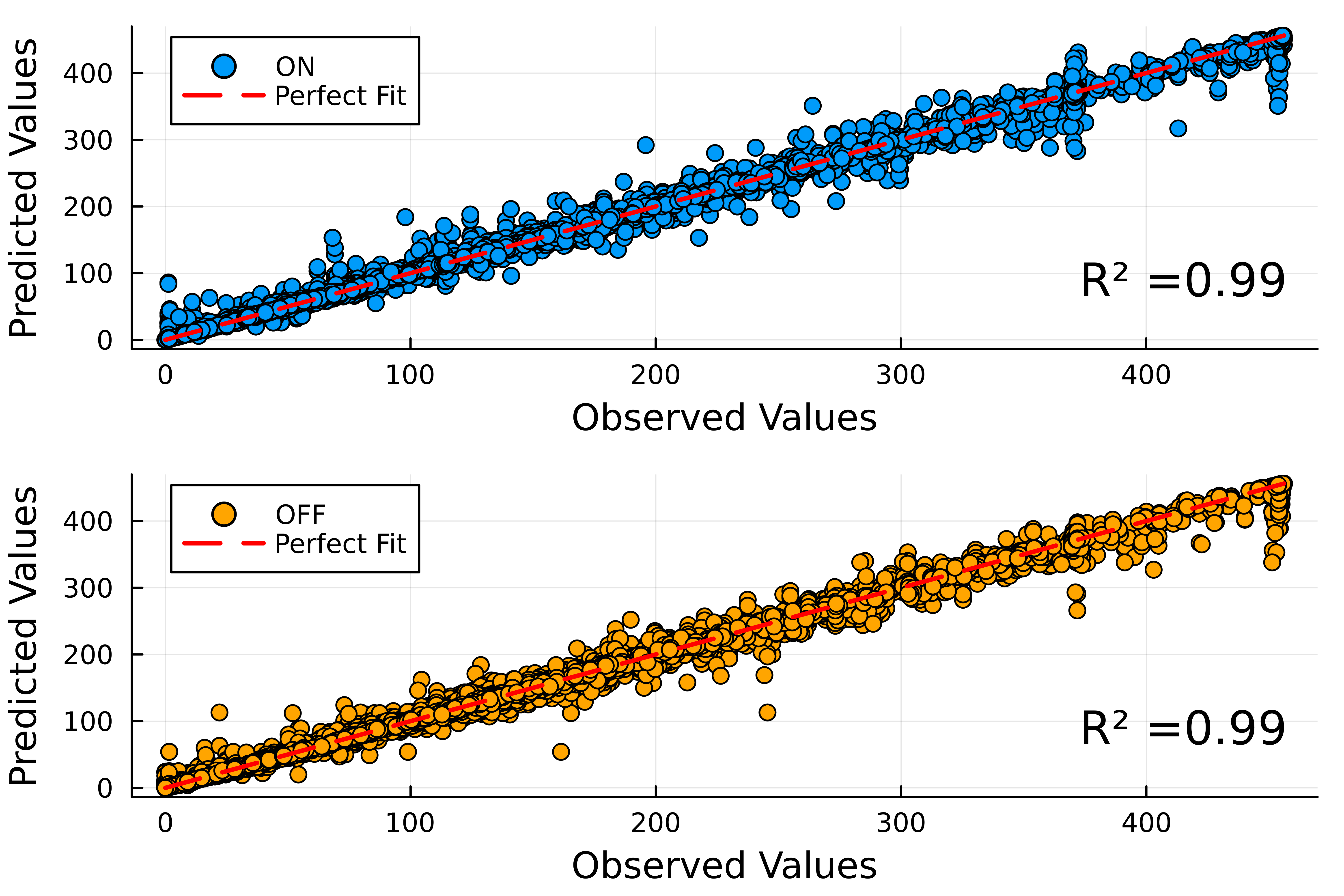}
\caption{}
\end{subfigure}
\begin{subfigure}[b]{0.7\textwidth}
\centering
\includegraphics [width=7.5cm] {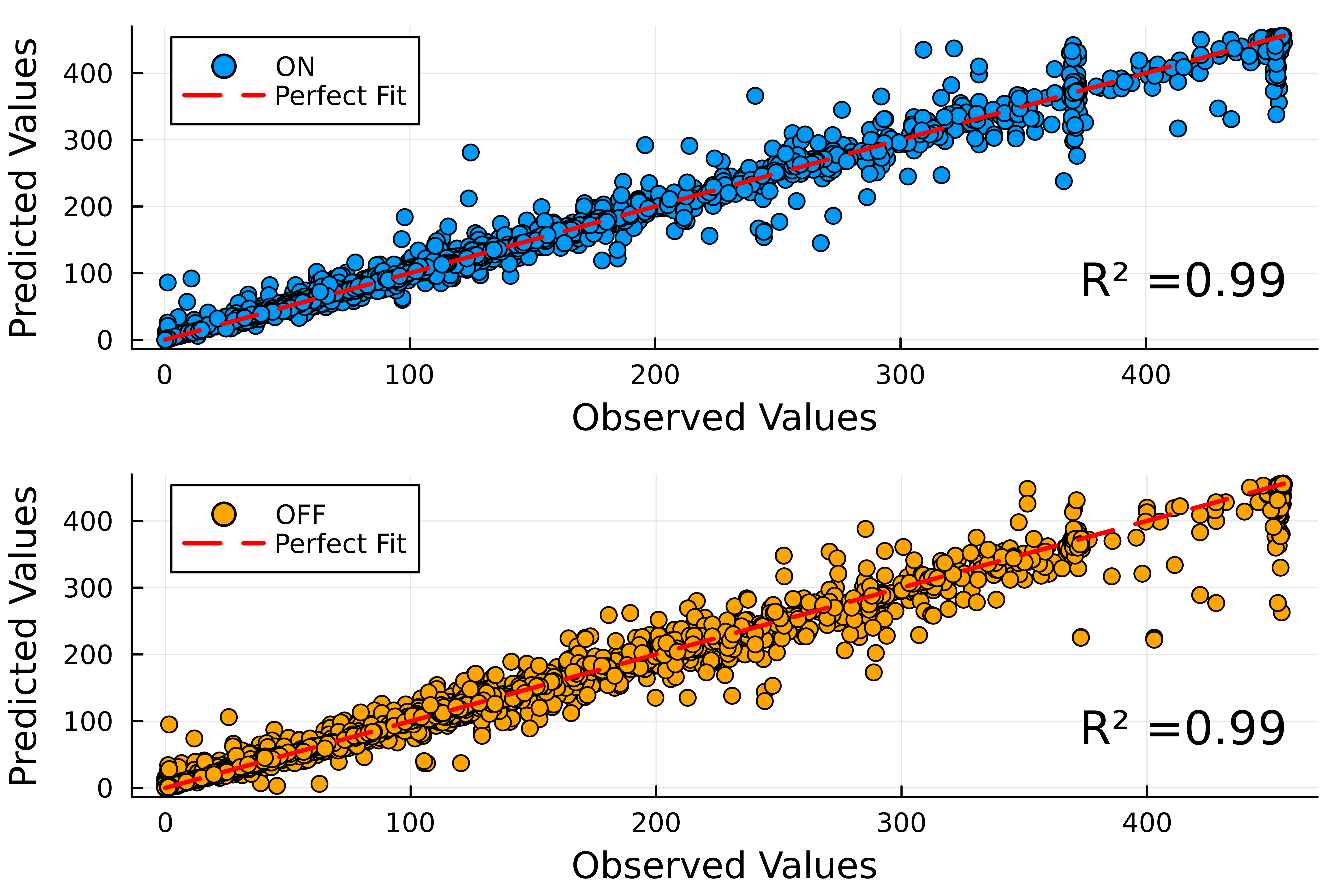}
\caption{}
\end{subfigure}
\caption[Model evaluation]{a) Evaluation of the machine learning model on the training data. b) Evaluation of the model on the testing data. the diagonal dashed line represent where datapoints should fall if there was a perfect match between experimental data and model.}
\label{fig:ML_eval}
\end{figure*}

\begin{figure*}[h]
\centering
\begin{subfigure}[b]{0.7\textwidth}
\centering
\includegraphics [width=7.5cm] {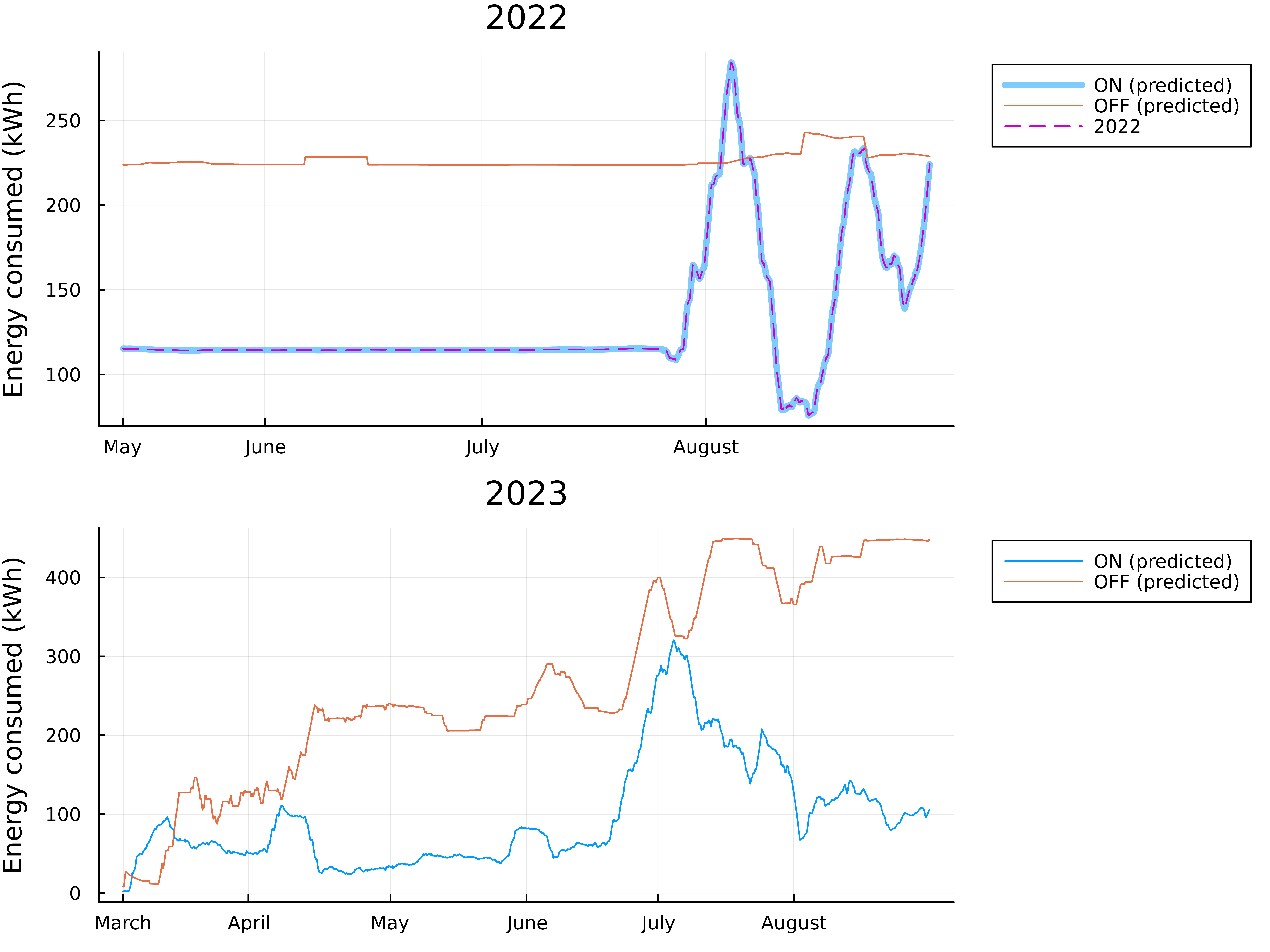}
\caption{}
\end{subfigure}
\begin{subfigure}[b]{0.7\textwidth}
\centering
\includegraphics [width=7.5cm] {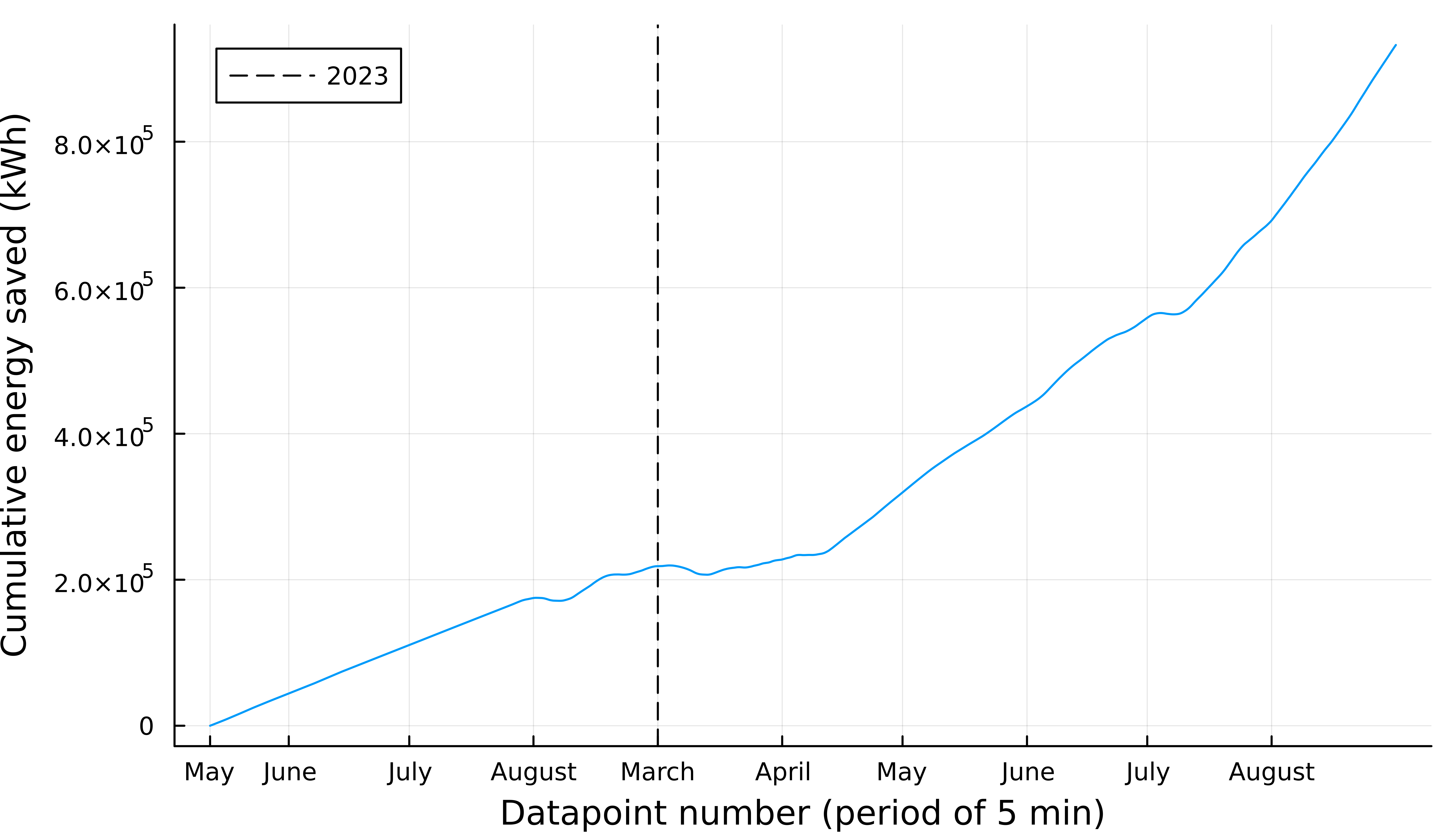}
\caption{}
\end{subfigure}
\caption[Energy saved]{a) Predicted energy saved in 2022 and 2023 with the filter always operating for periods of 5 minutes. b) Predicted cumulative energy saved between 2022 and 2023 with the filter always operating. The vertical dashed line represents the rupture in the dataset between September 2022 and March 2023.}
\label{fig:ML_saved}
\end{figure*}

The machine learning model was trained 
to account for both investigated configurations of the filter (Figure \ref{fig:ML_eval}). The model exhibited an R-squared value of 0.99 between the predicted and observed values across both the training and testing datasets. This coefficient of determination implies that 99\% of the variance in the predictions was accurately captured by the model, showcasing significant predictive ability. This trained model was then compared to the available data. Because the filter was operated throughout the 2022 cooling season, the electric energy consumed, as well as information on all the necessary features, was available. As can be observed in figure \ref{fig:ML_saved} a), the prediction of the model for the electric energy consumed by the system almost perfectly matched the observed values, further indicating the reliability of the model. Indeed, the line for the 2022 data with the filter operating and the line for the predicted data with the filter ON are superposed (Figure \ref{fig:ML_saved} a). 

The predicted electric energy consumed by the system during 2022 showed little variability between May and August for both configurations, but the configuration with the filter operating displayed a standard deviation an order of magnitude higher during the month of August (Figure \ref{fig:ML_saved} a). Nonetheless, if the filter had been OFF in 2022, the system would have consumed 218 000 more kWh, equivalent to 5\% of the total consumed energy. As the year 2023 featured record high temperatures, causing higher thermal loads during the cooling season \cite{zheng2023will}, the average predicted consumed electric energy was 26\% higher (Figure \ref{fig:ML_saved} a). The model predicted that the system would have saved 714 000 kWh between March and September 2023, equivalent to 13\% of the consumed energy, with the filter operating. Cumulatively, the energy saved increased linearly between May and September 2022 and during the cooling season in 2023, from May to September (Figure \ref{fig:ML_saved} b). Assuming the typical industrial energy cost of 0.07 USD/kWh in North America for large-power consumers \cite{quebec2023comparison}, 15 265 USD would have been saved after 2022 and a cumulative total of 62 530 USD would have been saved after 2023. 
\par

\par 
As can be observed in figure \ref{fig:ML_saved}, the energy saved was most significant during the main cooling season (between May and September). The difference between the consumed energy for each configuration was not as significant between March and mid-April, as the cumulative saved energy only increased by 5\% during that period. It can thus be concluded that during the colder months, between September and March, energy savings would be marginal. As was also observed in figure \ref{fig:COP_expl}, the operation of the filter during periods of lower thermal loads and wet-bulb temperatures sometimes led to slightly more inefficient heat transfer when the filter was operating. Therefore, it might be more costly to operate the system with the filter ON during that period. Nevertheless, it is crucial to emphasize that the operation of the filter throughout the year is instrumental in achieving energy savings during the cooling season, which coincides with the peak of evaporative cooling activity. Maintaining continuous filter operation throughout the year significantly mitigates biofouling on the surfaces of the cooling tower system, thereby improving evaporative cooling efficiency, suggesting potential long-term reductions in maintenance costs \cite{EvoquaBiofouling}. Thus, while the benefits mostly manifest during the cooling season (high thermal loads), they are contingent upon year-round filter operation. Moreover, the benefits would be extended in geographical regions where the cooling season is longer.

\subsection{Effect of intermittent operation}

\begin{figure*}[h]
\centering
\begin{subfigure}[b]{0.7\textwidth}
\centering
\includegraphics [width=7.5cm] {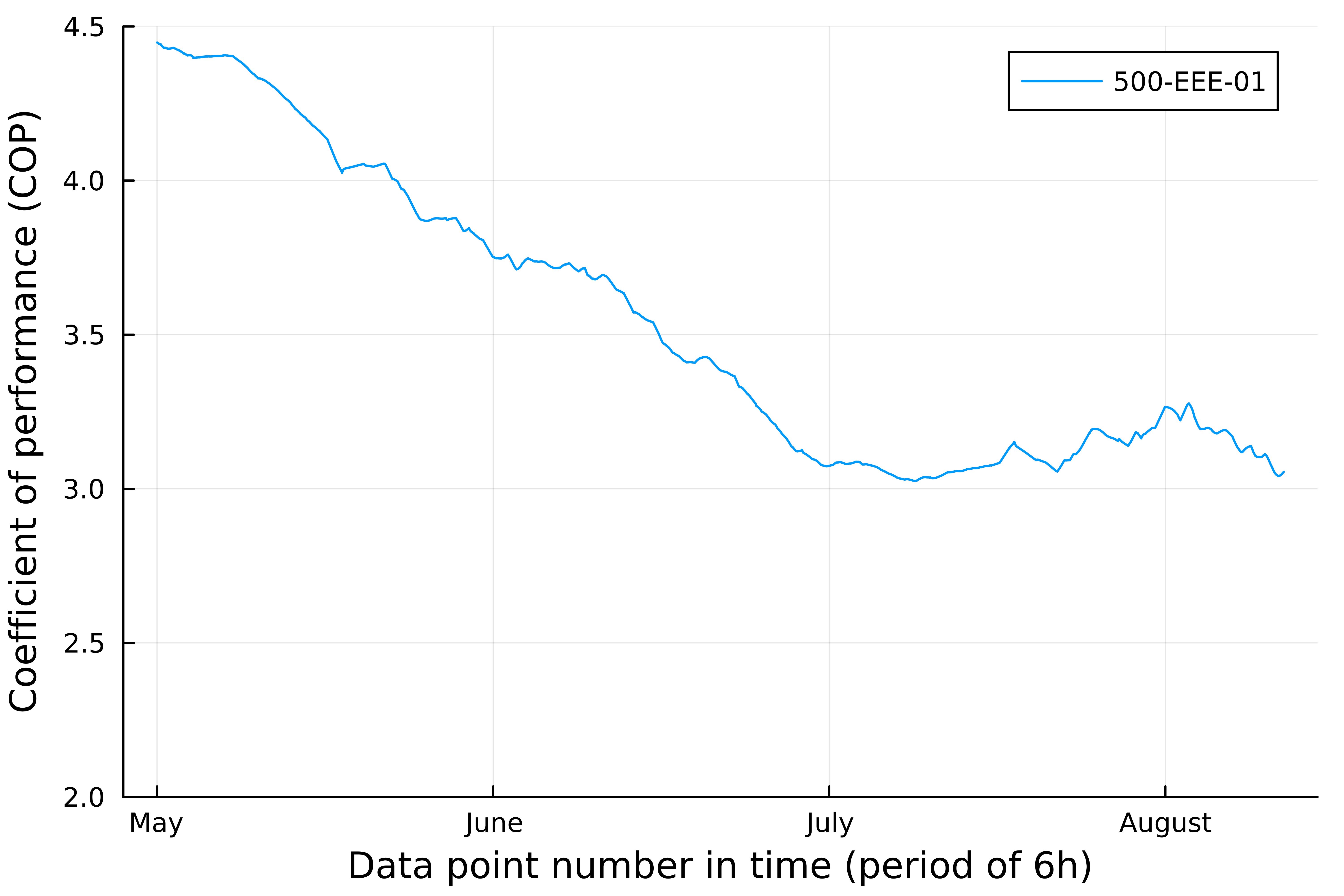}
\caption{}
\end{subfigure}
\begin{subfigure}[b]{0.7\textwidth}
\centering
\includegraphics [width=7.5cm] {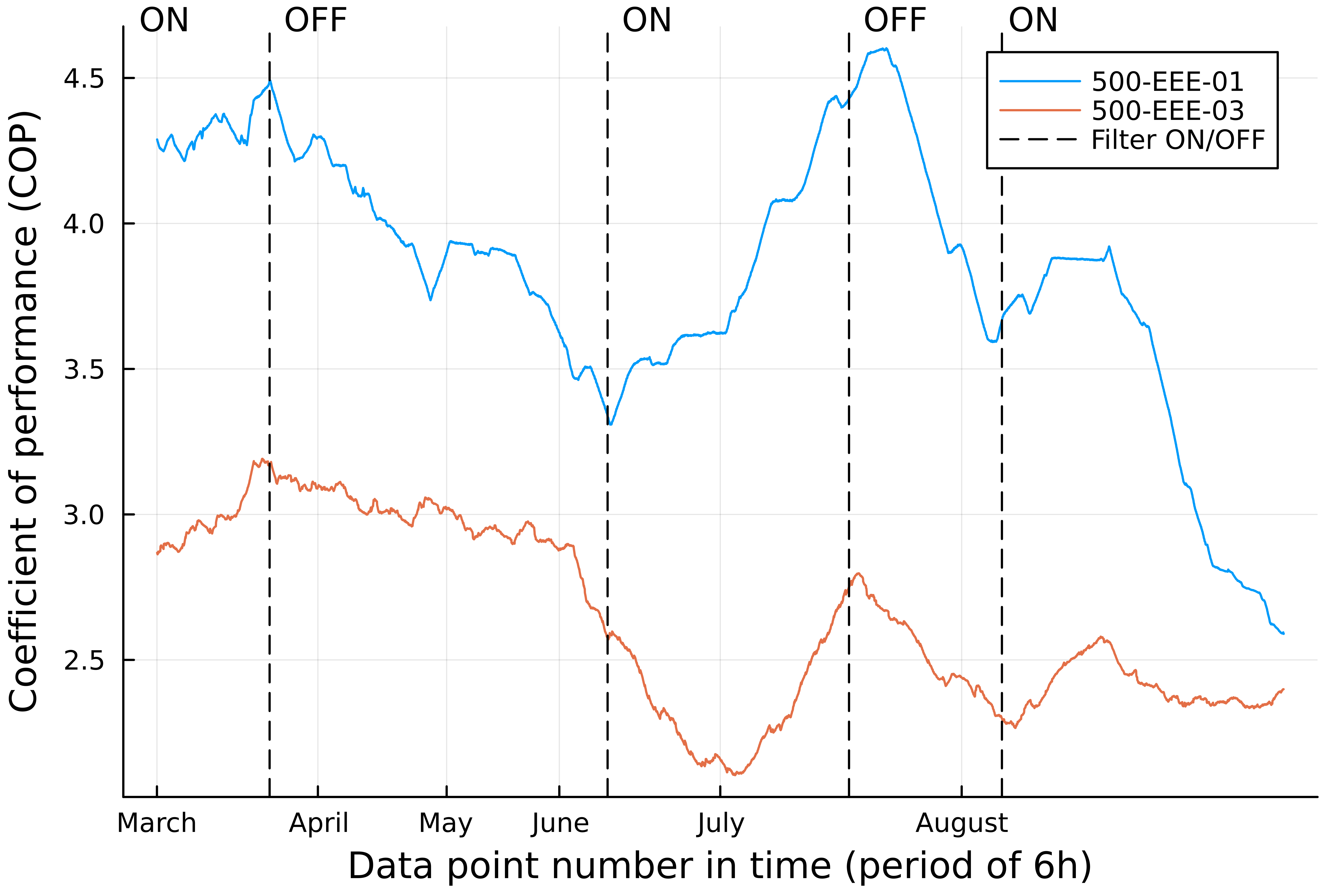}
\caption{}
\end{subfigure}
\caption[Subsystem analysis]{a) Behavior of heat exchanger 1 in 2022. b) Behavior of heat exchanger 1 and 2 in 2023.}
\label{fig:subsystem}
\end{figure*}

The intermittent operating pattern of the filter on the coefficient of performance of each of the heat exchangers was investigated (Figure \ref{fig:subsystem}). At the beginning of the cooling season in May 2022, the coefficient of performance of heat exchanger 1 was 4.5, its highest during that year. Similar results were also observed at the beginning of March 2023, as well as with the coefficient of performance of heat exchanger 2, albeit slightly lower at approximately 3. Thus, the highest coefficient of performance observed was at the beginning of the cooling season during both years of the investigation. In 2022, the performance of heat exchanger 1 decreased slightly during the cooling season to approximately 3, equivalent to a 50\% decrease in performance, even though the filter was always operating. This phenomenon of steady decrease in performance throughout the cooling season was also observed for both heat exchangers during the 2023 cooling season with a decrease of 50\% in performance for heat exchanger 1 and 25\% for heat exchanger 2 between March and September. It is typical for the performance of heat exchangers to decrease up to 50\% during the cooling season because of the biofilm accumulation, as well as because of the increased load on the system, which can be unevenly distributed between the subsystems \cite{al1997optimum}. As such, the coefficient of performance of the subsystems tends to decrease even with the filter operating. 
\par

When the operation of the filter was first stopped at the end of March, the coefficient of performance started to decrease linearly almost immediately for both heat exchangers. A decrease of 28\% and 20\% was observed within 8 weeks for heat exchangers 1 and 2 respectively. When the operation of the filter was resumed in June, the coefficient of performance of both heat exchangers increased to levels almost as high as at the beginning of the year. Recovery time was approximately 4 weeks, just before the second filter shutdown. A decrease of 25\% and 16\% in the coefficient of performance for heat exchangers 1 and 2 respectively was observed again when the filter stopped operating a second time in July. The decrease of the coefficient of performance was almost three times faster (steeper slope) during the second shutdown, likely because the wet-bulb temperature was higher and biofilm could develop faster. Furthermore, while the filter was operated between September 2022 and March 2023, the coefficient of performance of heat exchanger 1 increased from approximately 3 to 4.5, indicating that when the filter is operated during periods of lower thermal load, the coefficient of performance of the subsystems tends to increase. The performance of the heat exchangers increased linearly when the filter was operating and decreased linearly when it was not (Figure \ref{fig:subsystem} b). This suggests that operating the filter slows down the decrease in the coefficient of performance of individual subsystems during the cooling season.
\par
These results indicate that the filter has a tangible effect on the biofouling inside the heat exchangers as well as throughout the cooling tower system. The effect of biofouling on the performance of a cooling tower system can impede the heat transfer capacity by up to 50\% \cite{paniagua2020impact, microorganisms9030577}. As the performance of the heat exchangers decreased by up to 28\% when the filter was not operating, the biofilm formation likely increased significantly after the filter was stopped \cite{EvoquaBiofouling}. After resuming the operation of the filter, the observed linear increase in the coefficient of performance was likely due to the cleaning effect of the filter on the cooling tower water in the system. 
These results further emphasize the importance of operating the filter all year. Filtering cooling tower water in conjunction with standard chemical water treatment appears to offer a notable advantage by substantially mitigating the buildup of biofilm, increasing the evaporative cooling efficiency of the cooling tower system while concurrently minimizing the risks of bacterial contamination. 

\section{Conclusion}
A data-driven comprehensive analysis was conducted over two cooling seasons (2022 and 2023) on the energetic performance of a cooling tower system using a cross-flow microsand filtering system. The results underscore the intricate relationship between wet-bulb temperature, thermal load and cooling tower performance. As wet-bulb temperature and thermal load rise, the efficiency of cooling tower systems tends to decrease. Filtration emerges as an energy and cost-effective solution that enhances cooling tower performance. A time series analysis demonstrated that the coefficient of performance was on average 18\% higher, was higher 63\% of the time when the filter was operating and for 81\% of the total transferred energy, most markedly at higher thermal loads. These results reveal that over time, the cooling tower system was more efficient at conducting heat transfer when using filtration. Furthermore, during periods of high cooling demand characterized by increased wet-bulb temperatures and thermal loads, the performance gap widened significantly, with the system operating 41\% more efficiently with the filter in operation. Predictions for the energy saved over time with the filter operating were conducted using a machine learning model and indicated that 5 and 13\% of the energy bill could have been saved during the 2022 and 2023 cooling seasons respectively by operating the filter all year long, including the periods where the operation of the filter is more energetically costly. Notably, most energy savings occur during the cooling season, thus enhancing overall system durability and minimizing environmental impact. In addition, the biofouling mitigation provided by continuous filter operation \cite{EvoquaBiofouling} enabled higher performance of the system, underscoring the importance of continuous filter operation. Integrating cross-flow microsand filtration systems into cooling tower management not only optimizes performance but also aligns with sustainable practices, mitigating microbial risks and conserving energy resources.

\section*{Acknowledgements}
This research was supported by the MITACS Accelerate program with Evoqua+Xylem industrial partner (project Mitacs IT27607). Xavier Lefebvre is supported by the Hydro-Québec excellence scholarship as well as a NSERC scolarship (ES D) with a FRQNT supplement (B2X).
\par
We want to acknowledge the contribution of Alain Silverwood, Technical Director R\&D at Xylem / Evoqua Water Technologies, this project would not have been possible without his vision, his continuous support for experiment design, data collection, field support and training. 
\par
Special Thanks to the study site managers for access to their building and data.
\par
Thank you also to Nils Frejinger Robert for his contribution to the time series algorithm that was implemented into the code. 

\section*{Conflict of interest}
The authors declare no competing interests.

\section*{Supporting Information}
 The data that support the findings of this study are available upon reasonable
request from the authors.

\printendnotes

\bibliography{references}

\end{document}